\documentclass[multphys,vecphys]{svmult}

\usepackage{makeidx}         % allows index generation
\usepackage{graphicx}        % standard LaTeX graphics tool
\usepackage{psfrag}
\usepackage{subfigure}
\usepackage{cite}                     % 
\usepackage{multicol}        % used for the two-column index
\usepackage[bottom]{footmisc}% places footnotes at page bottom
\makeindex             % used for the subject index
\newcommand{\Eq}[1]{{(\ref{#1})}}

\begin{document}

\title*{Aspects of the FM Kondo Model: From Unbiased MC
  Simulations to Back-of-an-Envelope Explanations}
\titlerunning{Aspects of the FM Kondo Model} 
\author{Maria Daghofer\inst{1}\and
Winfried Koller\inst{2}\and Alexander Pr\"ull\inst{1}\and Hans Gerd
Evertz\inst{1}\and Wolfgang von der Linden\inst{1}}
\authorrunning{Daghofer et al.}

\institute{Institute for Theoretical and Computational Physics\\
   Graz University of Technology\\
\texttt{daghofer@itp.tu-graz.ac.at}
\and Department of Mathematics, Imperial College}
\maketitle

\begin{abstract}
Effective models are derived from the ferromagnetic Kondo lattice
model with classical corespins, which greatly reduce the numerical
effort. Results for these models are presented. They indicate that
double exchange gives the correct order of magnitude and the correct
doping dependence of the Curie temperature. Furthermore, we find that
the jump in the particle density previously interpreted as phase
separation is rather explained by ferromagnetic polarons.
\end{abstract}

% \section{Introduction}
% \label{sec:Intro}

Manganites~\cite{proceedings98} are often described by the ferromagnetic Kondo lattice model, which is
considered to explain some of their features, e.g., the
transition from antiferromagnetic to ferromagnetic order with doping~\cite{dagotto01:review}.

The application of the model is motivated by the fact, that
crystal field splitting divides the five d-orbitals into two $e_g$  and three
$t_{2g}$ orbitals, where the latter are energetically favored in the case of
manganites. All three $t_{2g}$ orbitals are singly occupied and rather
localized. Due to a strong Hund's rule coupling, these electrons are
aligned in parallel and form a core spin with length $S =3/2$. The
filling of the $e_g$ orbitals is determined by doping and these 
electrons can hop from one Mn ion to the next via the intermediate oxygen.
Hund's rule coupling leads to a
ferromagnetic interaction between the itinerant $e_g$ electrons
and the $t_{2g}$ core spin.
The core spins interact through
super exchange leading to a weak antiferromagnetic coupling between
them. 

In this chapter, we derive effective models for the
ferromagnetic Kondo lattice model and introduce suitable Markov chain Monte
Carlo (MC) algorithms. The presented results, were not obtainable by simple
analytic considerations, are partly found by this MC method and partly by
use of the Wang-Landau algorithm~\cite{wang01:_deter}. 

\section{Model Hamiltonian}
\label{sec:Model_Hamiltonian}

The spin and charge degrees of freedom in manganites can be described by the ferromagnetic Kondo lattice model with two
orbitals ($x^2-y^2$ and $3z^2 - r^2$):
\begin{equation}\label{def:FKLM}
    \hat H = -\sum_{i,j,\alpha,\beta, \sigma}\;
  t_{i\alpha,j\beta}\;
  c^\dagger_{i\alpha\sigma}\,c^{\phantom{\dagger}}_{j\beta\sigma}
  - \tilde J_H \sum_{i\alpha} \vec{\sigma}_{i\alpha} \cdot \vec{S}_i 
  + J'\sum_{<ij>} \vec{S}_i \cdot \vec{S}_j\;,
\end{equation}
where $c^\dagger_{i\alpha\sigma}$
($c^{\phantom{\dagger}}_{i\alpha\sigma}$) creates (annihilates) an electron with spin
$\sigma$ in orbital $\alpha$ at site $i$, $\vec{\sigma}_{i\alpha}$ denotes
electron spin in orbital $\alpha$ and $\vec{S}_i$ the core spin.
The first term of the Hamiltonian describes the hopping between the
nearest neighbor
sites; the hopping strength $t_{i\alpha,j\beta}$ depends on the
involved orbitals and the direction. 
As matrices in the orbital indices $\alpha,\beta=1 (2)$, corresponding to the $x^2-y^2$
($3z^2-r^2$) orbitals (see e.g. \cite{dagotto01:review}), this hopping reads
\begin{equation}\label{eq:hopping}
  t_{i,i+\hat z} \;=\; t \;
    \left(\begin{array}{cc}
    0 & 0 \\ 0 & 1
    \end{array} \right),\quad
  t_{i,i+\hat x/\hat y} =  t \;
    \left(\begin{array}{cc}
    3/4 &\mp\sqrt{3}/4 \\ \mp\sqrt{3}/4 & 1/4
    \end{array} \right)\;.
\end{equation}
The overall hopping strength is $t$, which
will be used as unit of energy, by setting $t=1$.

The second term contains the
ferromagnetic interaction between the electrons and the core spins and
the third term is the AFM superexchange of the core spins.

The $S =3/2$ core spin can approximately be treated as a
classical spin, which corresponds to the limit $S \rightarrow
\infty$~\cite{furukawa98,dagotto98:_ferrom_kondo_model_mangan}. It is then
replaced by a vector of unit length and the factor 
$3/2$ is incorporated into $\tilde J_H$. This approximation simplifies
calculations enormously and should not lead to much difference from the
quantum case except possibly for very low temperatures 
$T\approx 0$~\cite{Nolting01,Nolting03,dagotto98:_ferrom_kondo_model_mangan,
Garcia_FMPOL02,NeuberDaghofer2005}.

For large $\tilde J_H$, the electronic density of states of the Hamiltonian \Eq{def:FKLM}
is split into the lower and upper Kondo band, separated by
approximately $\tilde J_H$. The $e_g$ electrons move mostly
parallel to the core spins in the lower band, and anti-parallel in the upper band.
 
In order to derive effective low-energy models for $\tilde J_H \gg t, J'$, we
change the quantization
axis for the electron spin from the \emph{global} quantization axis
(e.g. the z-direction) to a \emph{local} quantization axis, namely
the direction of the local $t_{2g}$ core spin. Spin up (down)  then means that the $e_g$
electron spin is parallel (antiparallel) to the core spin. The Hund's
rule term $\tilde J_H\; \vec{\sigma}_{i\alpha}\cdot\vec{S}_i$ becomes $J_H\;
(\hat n_{i\alpha\downarrow} - \hat n_{i\alpha\uparrow})$, with the
factor $1/2$ coming from the electron spin also absorbed into
$J_H$. While the
$e_g$-spin is preserved in global quantization, this is no longer the case in the
local quantization. An up electron at site $i$ can therefore become a
down electron at site $j$, which is denoted by the superscript for the
hopping strength. Furthermore, the hopping now depends on the core spins:
\begin{equation}\label{eq:hopp_u}
t^{\sigma,\sigma'}_{i\alpha,j\beta} = t_{i\alpha,j\beta} u^{\sigma,\sigma'}_{ij}
\end{equation}
The first factor is the orbital-dependent hopping strength
\Eq{eq:hopping} and the second factor contains the relative orientation of the
core spins:
\begin{eqnarray}
% \begin{aligned}
    u^{\sigma,\sigma}_{i,j} =& c_i c_j + s_i s_j \;
    \E^{i\sigma (\phi_j-\phi_i)}&=
    \cos(\vartheta_{ij}/2)\;\E^{i\psi_{ij}}\\
    u^{\sigma,-\sigma}_{i,j} =& \sigma(c_i s_j\;e^{-i \sigma \phi_j} -
    c_j s_i \; e^{-i\sigma \phi_i})&=
    \sin(\vartheta_{ij}/2)\;\E^{i\chi_{ij}}
%  \end{aligned}\;,
\end{eqnarray}
with the abbreviations $c_j = \cos(\theta_j/2)$ and $s_j = \sin(\theta_j/2)$ and
the restriction $0\le\theta_j\le\pi$, where $\theta_i, \phi_i$ are the
polar coordinates for core spin $\vec S_i$. These factors depend on the
relative angle $\vartheta_{ij}$ of the
core spins $\vec S_i$ and $\vec S_j$ and on some complex phases
$\psi_{ij}$ and $\chi_{ij}$.
With a shift of the chemical potential $\mu \rightarrow \mu - J_H$,
the Hamiltonian  \Eq{def:FKLM} in local spin-quantization reads:
\begin{equation}\label{FKLM_loc}
  \hat H = -\sum_{i,j,\alpha,\beta, \sigma,\sigma'}\;
  t^{\sigma,\sigma'}_{i\alpha,j\beta}\;
  c^\dagger_{i\alpha\sigma}\,c^{\phantom{\dagger}}_{j\beta\sigma'}
  + 2 J_H \sum_{i\alpha} \hat{n}_{i\alpha\downarrow} %
  + J'\sum_{<ij>} \vec{S}_i \cdot \vec{S}_j\;.
\end{equation}
This is still the same Hamiltonian as \Eq{def:FKLM} without any approximation besides
the use of classical core spins.

\subsection{Effective Spinless Fermions}
\label{sec:ESF}

Most of the experimental results on manganites and all of the
theoretical work presented here concerns electron densities $0 \leq
n_\textnormal{el} \leq 1$, i.~e. predominantly the lower Kondo
band. As $J_H$ is much larger than the hopping $t$ and the AFM
superexchange $J'$, one can simplify the model by a separation of
energy scales \cite{Auerbach:book}. As a first approximation, one can
take $J_H \rightarrow \infty$ and thereby leave out the configurations
with $e_g$ electrons antiparallel to the core spins completely. This
approximation is widely used \cite{Aliaga_island_2d,Yamanaka_98}, but misses some
important effects discussed in Sect. \ref{sec:val_esf_uha}. However, if one treats these configurations in second order
perturbation theory \cite{yarlagadda01:mf_COSO,SQShen00:pp_mf}, almost
perfect agreement to the original Kondo lattice model is obtained without any
additional numerical effort \cite{KollerPruell2002a}. This approach is 
similar to the derivation of the $t-J$ model from the Hubbard model,
while the $J_H \rightarrow \infty$ method corresponds to $U\rightarrow
\infty$ for the Hubbard model.

In this effective model, the dynamical degrees of freedom are the low energy states with the
$e_g$-spins parallel (i.~e. up in local quantization) to the $t_{2g}$-spins. The virtual excitations
meditated by the hopping matrix are configurations where one $e_g$
electron is antiparallel (down):\\
$(i\alpha\uparrow) \to
(j\beta\downarrow) \to (i'\alpha'\uparrow)$. As the low energy states
contain only up electrons, this lead to an effective spinless fermion
Hamiltonian:
\begin{equation}                                       \label{def:ESF}
  \begin{array}{rcl}
  \hat H_\textrm{ESF} = &-&\sum\limits_{i,j,\alpha,\beta}
    t^{\uparrow\uparrow}_{i\alpha,j\beta}\,
    c^\dagger_{i\alpha}\,c^{\phantom{\dagger}}_{j\beta}
    - \sum\limits_{i,\alpha,\alpha'}
    \bigg(\sum\limits_{j,\beta}
    \frac{t^{\uparrow\downarrow}_{i\alpha',j\beta}\,t^{\downarrow\uparrow}_{j\beta,i\alpha}}
    {2J_H}\bigg) c^\dagger_{i\alpha'}c^{\phantom{\dagger}}_{i\alpha}\\
    &-& \sum\limits_{[i\ne i'],\alpha,\alpha'}\!\!\!
    \bigg(\sum\limits_{j,\beta}
    \frac{t^{\uparrow\downarrow}_{i'\alpha',j\beta}\,t^{\downarrow\uparrow}_{j\beta,i\alpha}}
    {2J_H}\bigg) c^\dagger_{i'\alpha'}c^{\phantom{\dagger}}_{i\alpha}
    + J'\sum\limits_{<ij>} \vec{S}_i \cdot \vec{S}_j \;.
  \end{array}
\end{equation}
The first term of this Hamiltonian contains the kinetic energy of the
electrons moving in the lower Kondo band. As
$t^{\uparrow\uparrow}_{i\alpha,j\beta}$
is largest for parallel core spins, this term favors
ferromagnetism. The second term describes electrons that get excited
into the upper Kondo band and then hop back to the original site. It
yields a density dependent antiferromagnetic interaction between the
core spins. The third term is a `three-site-term' of minor
influence and will be neglected~\cite{KollerPruell2002a}. On the other hand, its inclusion does not increase the
numerical effort.

The reduction of the Hilbert space achieved by this effective model is
the same as for the $J_H \rightarrow \infty$ limit, and finite $J_H$
can thus be treated with the same numerical effort.

\subsection{Uniform Hopping Approach}
\label{sec:UHA}

A significant further simplification is the uniform hopping
approximation proposed by van den Brink and Khomskii
\cite{brink99:_DE_two_orbital}. This approximation replaces the
different angles of neighboring core spins by a mean value. In order to
treat anisotropies, two different
angles are chosen, $\theta_z=\vec{S}_i\cdot\vec{S}_{i\pm z}$ in
z-direction and $\theta_{xy}=\vec{S}_i\cdot\vec{S}_{i\pm x}=\vec{S}_i\cdot\vec{S}_{i\pm y}$ within the
$xy$-plane. These should not be confused with the polar angle of an
individual core spin $\theta_i$. It is assumed that the relative orientation is the same
between all nearest neighbor pairs. The hopping matrix therefore becomes 
translationally invariant. Spin configurations that are still treated exactly include,
among others, ferro- and antiferromagnetism and spin canted states.

The impact of the core spins on the hopping simplifies to
\begin{equation}
  u^{\sigma,\sigma}_z  = \cos(\frac{\theta_z}{2}) = u_z\quad,\qquad
  u^{\sigma,-\sigma}_z = \sin(\frac{\theta_z}{2}) = \sqrt{1-u_z^2}\
\end{equation}
in z-direction and analogously in x/y-direction. Likewise, the inner product of
the $t_{2g}$ spins entering the superexchange term can be expressed by
\begin{equation}
  \mathbf S_i\cdot \mathbf S_{i+\hat z} = \cos\theta_z = 2u_z^2-1\;.
\end{equation}

The energy of this model can easily be evaluated, especially in the
thermodynamic limit and the ground state can be obtained by minimizing
the energy with respect to $\theta_z$ and $\theta_{xy}$.

For a one orbital model in one dimension with
periodic boundary conditions, the Hamiltonian simplifies to
\begin{equation}\label{UHA_Ham_1d}
      \hat H = -u_z \sum_{\langle i j\rangle} c^\dagger_{i}c^{\phantom{\dagger}}_{j}
  - \frac{1-u_z^2}{J_H}\,\sum_i c^\dagger_{i}c^{\phantom{\dagger}}_{i}
  +J'L\big(2u_z^2-1\big)\;,
\end{equation}
This Hamiltonian yields a shifted tight-binding band structure
\begin{equation}\label{eq:dispersion}
    \epsilon_k = -2u\cos(k) - (1-u_z^2)/J_H
\end{equation}
with a band width of $4\;u_z$, which has its maximum for
ferromagnetic core spins and vanishes for antiferromagnetic order.

Similar calculations can be done for three dimensions with both
orbitals and a ground state phase diagram can thus be obtained, see \cite{brink99:_DE_two_orbital,KollerPruell2002a}.

This UHA approach was extended to finite temperatures in \cite{KollerPruell2002b}. To
introduce this method, we proceed as follows:
For a given core spin configuration $\mathcal S$, characterized by the set of
angles $\{\theta_i,\phi_i\}$, we define the average $u$-value
\begin{equation}
  u(\mathcal S) = \frac{1}{N_p} \sum_{\langle i j\rangle}
  u^{\uparrow\uparrow}_{ij}(\mathcal S)\;.
\end{equation}
Here $N_p$ is the number of n.n.\ pairs $\langle i j\rangle$.
In the ESF Hamiltonian \Eq{def:ESF}, $u^{\uparrow\uparrow}_{ij}$ is
then replaced by $u(\mathcal S)$.
Besides $u^{\uparrow\uparrow}_{ij}$ the Hamiltonian depends on
$|u^{\sigma,-\sigma}_{ij}|^2$ and on $\mathbf S_i\cdot\mathbf S_j$,
which correspond to $\sin^2(\vartheta_{ij}/2)$ and $\cos\vartheta_{ij}$,
respectively.
As a further approximation, these terms are replaced
by $1-u^2(\mathcal S)$ and $2\,u^2(\mathcal S)-1$ respectively, which leads to the
one-orbital UHA Hamiltonian
\begin{equation}\label{UHA_Ham}
      \hat H = -u \sum_{\langle i j\rangle} c^\dagger_{i}c^{\phantom{\dagger}}_{j}
  - \frac{1-u^2}{2 J_H}\,\sum_i z_i c^\dagger_{i}c^{\phantom{\dagger}}_{i}
  +J'N_p\big(2u^2-1\big)\;,
\end{equation}
with $z_i$ being the number of nearest neighbors for site $i$. This
Hamiltonian defines the Boltzmann factor for the spin configuration
$\mathcal S$. In order to calculate thermodynamical expectation
values, one still needs to calculate the density of states $\Gamma(u)$, i.e. the
number of spin configurations with the same average value for $u$. It
can be calculated exactly in one dimension and by use
of the Wang-Landau algorithm \cite{wang01:_deter} in higher
dimensions. 

Once the density of states $\Gamma(u)$ has been obtained, observables can
be obtained for any temperature, much larger lattices can be treated
and a 3D phase diagram for finite temperatures can be obtained.
The numerical effort is
reduced from an integration over the $L$-dimensional space
of the core spin configurations to an integral over the one-dimensional
unit interval for $u$.

\section{Monte Carlo Algorithm}
\label{sec:alg}

The algorithm used to simulate the Kondo lattice model and the
effective spinless fermion model in the grand canonical ensemble is
the one proposed in \cite{yunoki98:_phase}. For each core spin
configuration, the resulting Hamiltonian for the $e_g$ electrons is a
one-particle problem. The
statistical weight for the core-spin configuration $\mathcal S$ in the
grand canonical ensemble is the starting
point of grand canonical Markov chain Monte Carlo simulations:
\begin{equation}                                  \label{eq:MC_weight}
    w(\mathcal S|\mu) = \frac{\textrm{ tr}_c\, \E^{-\beta (\hat H(\mathcal S)-\mu
    \hat{N})}}{\cal Z(\mu)}\;.
\end{equation}
It is calculated from
the eigenvalues of $\hat H(\mathcal S)$ 
by use of free fermion formulae, which is denoted by the trace over the fermionic degrees of
freedom $\textrm{ tr}_c$. 

As some particle numbers are not stable in the grand
canonical ensemble, we developed a canonical algorithm. An exact
approach would mean calculating the Boltzmann weight for every
possible distribution of $N_\textrm{el}$ electrons on $L$ eigenvalues
and summing over these contributions. This is numerically too demanding. But for low
temperatures, only very few of these distributions actually contribute
to the partition function. They can be obtained by
filling $N_0 < N_\textrm{el}$ electrons into the $N_0$ lowest
eigenenergies and distributing only the remaining $N_\textrm{el} -
N_0 \approx 5$ electrons on the states around the Fermi-energy. The
canonical weight then becomes:
\begin{equation}  \label{eq:MC_weight_can}
      w(\mathcal S|N_{el}) = \frac{\sum_{\tilde{\mathcal{P}}}\, \E^{-\beta \hat
      H(\mathcal S,\tilde{\mathcal{P}}(N_{el}) )}}{{\cal Z}(N_{el})}\;,
\end{equation}
where $\tilde{\mathcal{P}}$ denotes these restricted permutations.
In order to decrease autocorrelations, particle fluctuations within a
set of 3 to 5 densities were allowed. The amount of core-spin
rotations was small for most updates in order to ensure high
acceptance, but occasionally a complete spin
flip was proposed. In one dimension, whole sections of the chain were
rotated at once. 50 to several hundreds of sweeps were skipped between
measurements. This ensured statistical independence for the 1D
calculations, in 2D, remaining autocorrelations were treated by
autocorrelation analysis.

While observables depending only on $\cal Z$ are independent of the
spin quantization (global/local), care must be taken when evaluating
e.g. the one particle Greens function, which in global quantization
can be written as
\begin{equation}                                          \label{eq:MC_GF2}
  \sum_\sigma
  \ll a^{\phantom{\dagger}}_{i \sigma}; a^\dagger_{j \sigma} \gg_\omega
  = \int {\cal D}[\mathcal S]\; w(\mathcal S|\mu) u_{ji}^{\uparrow\uparrow}(\mathcal S)
    \ll c^{\phantom{\dagger}}_{i}; c^\dagger_{j} \gg^\mathcal S_\omega\;,
\end{equation}
where $\ll c^{\phantom{\dagger}}_{i}; c^\dagger_{j} \gg^\mathcal S_\omega$ is the
Green's function in local spin quantization.
It can be expressed in terms of the
one-particle eigenvalues $\epsilon^{(\lambda)}$ and the corresponding
eigenvectors $\psi^{(\lambda)}$ of the Hamiltonian $\hat H(\mathcal S)$:
\begin{equation}
\ll c^{\phantom{\dagger}}_{i}; c^\dagger_{j}\gg^\mathcal S_\omega
= \sum_\lambda \; \frac{\psi^{(\lambda)}(i) \;\psi^{*(\lambda)}(j)}
{\omega - (\epsilon^{(\lambda)} -\mu) + i0^+ }
\end{equation}
It should be pointed out that the one-particle density of states (DOS)
is identical in global and local quantization; for details see \cite{KollerPruell2002b}.

\section{Results}

In the first subsection, we will demonstrate the validity of the
simplified effective models. We will then use the uniform hopping
approach (UHA) in Sect.~\ref{sec:T_uha} to determine the Curie temperature of the one-orbital
model in three dimensions. In Sect. \ref{sec:PS_FP}, we will present
results for the one- and two-dimensional model and we will show that
they do not indicate phase separation, but rather ferromagnetic polarons. A
phase diagram for the 2D model is given in Sect.~\ref{sec:phase_diagr}.

\label{sec:results}

\subsection{Validity of the ESF Model and the UHA in one Dimension}
\label{sec:val_esf_uha}

In this section, we will present results from unbiased Monte Carlo
Simulations for the ESF model in one dimension with one orbital and
open boundary conditions and we will compare them to results for the
full ferromagnetic Kondo lattice model and the $J_H \rightarrow
\infty$ approximation. Simulations were done for $L=20$ sites,
$\beta=50, J'=0.02$ and $J_H$ varying from 4 to 10.

\begin{figure}[!ht]
\centering{\includegraphics[width=0.6\textwidth]{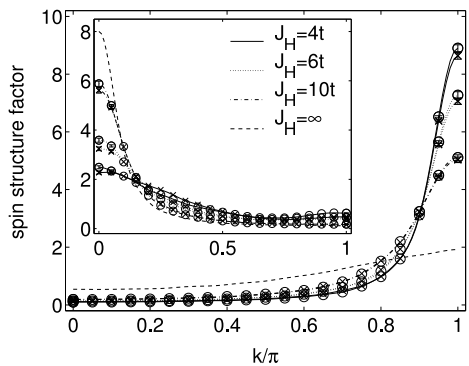}}\\[-0.2\textwidth]
\caption{Spin structure factor for the $t_{2g}$ spins at $n=1$ (inset:
  $n\approx 0.75$) for $\beta=50, J'=0.02, L=20$ and
  different values of $J_H$. Circles: spinless fermion model %$H_\textrm{ESF}$
  \Eq{def:ESF}; crosses: DE model \Eq{FKLM_loc}.
  In the limit $J_H\to \infty$ (dashed line) the intensity of the AFM peak is considerably
  smaller than for finite $J_H$. From Ref.~\cite{KollerPruell2002a}.
} \label{spin_structure_L20NxJx.eps}
\end{figure}

Figure \ref{spin_structure_L20NxJx.eps} shows the core-spin structure factor for
the three models. For electron density $n\approx 0.75$ (inset), the $t_{2g}$
correlations are ferromagnetic, driven by the
kinetic energy of the $e_g$ electrons. For $n=1$, the lower Kondo band
is completely filled, no hopping is possible and the kinetic energy
therefore vanishes. Excitations into the upper Kondo band (virtual for
the ESF), which are favored by antiferromagnetism, then dominate the energy.

For both densities and even for moderate $J_H=4$, the ESF model \Eq{def:ESF} and the original
Kondo Model \Eq{FKLM_loc} produce virtually identical results. The
$J_H \rightarrow \infty$ model on the other hand does not reproduce the antiferromagnetic
correlations for the completely filled lower Kondo band correctly, because
the virtual excitations are missing from this model, and it also
overestimates the ferromagnetic correlations at $n\approx 0.75$. 
The AFM effect coming from the virtual excitations can be described by
a \emph{density dependent} effective parameter
$J_\textrm{eff}=J' + 1/(2J_H)$ at $n=1$ and it is
generally much stronger than the small superexchange $J'$ also
favoring AFM. 

\begin{figure}[!ht]
  \centering{\includegraphics[width=0.5\textwidth]{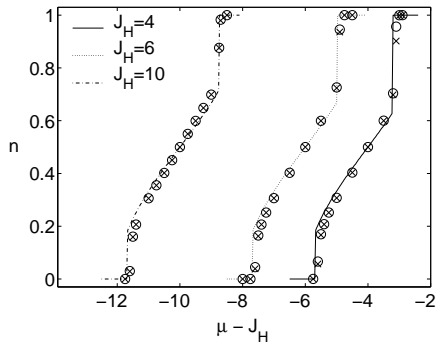}}
\caption{Electron density versus chemical potential for $J_H=4, 6$, and $10$ (right to left),
 and $J'=0.02$.
 MC results at  $\beta=50$, $L=20$ for the spinless fermion model
 $H_\textrm{ESF}$ (circles)
 are compared with those for the
 DE model $H$ (crosses). Error bars of the MC data are smaller than the symbols.
 The lines correspond to  groundstate UHA. From Ref.~\cite{KollerPruell2002a}.
}
\label{mu_n_L20Jx.eps}
\end{figure}

Figure \ref{mu_n_L20Jx.eps} shows the electron density versus the
chemical potential $\mu$ for the Kondo model, the effective spinless
fermions and groundstate UHA for $J_H =4, 6, 10$. All three models give
almost identical results. For very small $\mu$, the band is 
empty and $J'$ leads to antiferromagnetism. At a critical $\mu^c_1$
depending on $J_\textrm{eff}=J'$, the filling jumps to $n \approx 0.2$ and the
correlations become ferromagnetic. At a second critical  $\mu^c_2$,
depending on $J_\textrm{eff}=J' + 1/(2J_H)$, it becomes
antiferromagnetic again and the density jumps from $n \approx 0.7$ to
$n\approx 1$.

These discontinuities mean that intermediate particle numbers are not stable in
the grand canonical ensemble. They have been interpreted as phase
separation. However, we will show in Sect. \ref{sec:PS_FP} that their cause lies rather
in small ferromagnetic polarons.

\subsection{Finite Temperature UHA  and Curie Temperature}
\label{sec:T_uha}

Although the uniform hopping approach replaces the fluctuating
core spins by an average $u$, it reproduces not only the expectation
value  of the energy, but also its width with astonishing accuracy,
even for higher temperatures. The results of UHA also remain valid
upon the inclusion of n.n Coulomb repulsion~\cite{KollerPruell2002b}. 
All the while, the numerical effort is reduced
from sampling over hundreds of thousands of core-spin configurations to
scanning the single parameter $u$ within the unit interval. The fact
that the results remain valid with inclusion of the Coulomb repulsion
indicates that UHA is a reliable approximative method that can safely
be extended to more complicated situations.

\begin{figure}[h]
  \centering{\includegraphics[width=0.6\textwidth]{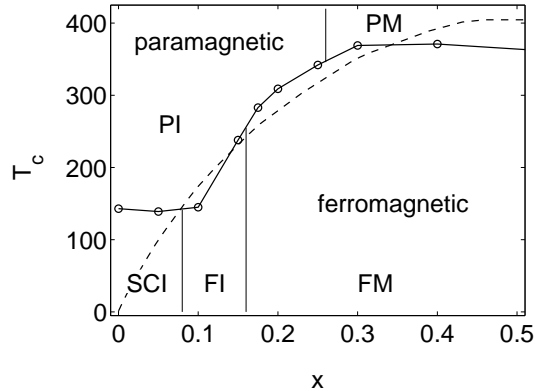}}
  \caption{Curie temperature (dashed line) of the one-orbital DE
    model for a $16^3$ cluster and $t=0.2\,$eV calculated in UHA.
    Circles and phases PM (paramagnetic metal), PI (paramagnetic
    insulator), 
    FM (ferromagnetic metal), FI (ferromagnetic insulator), and SCI
    (spin canted insulator) are experimental
    results for 
    La$_{1-x}$Sr$_{x}$MnO$_3$ \cite{urushibara95:_insul_la_sr_mno},
    UHA results \cite{KollerPruell2002b}.}
  \label{LaSrMnO_phases.eps}
\end{figure}

As observables can be evaluated for all temperatures once the density
of states $\Gamma(u)$ is known,
it can be used to determine the Curie temperature for the three
dimensional one-orbital model with $J_H=\infty$ and $J'=0$. If one
sets the only free parameter, namely the hopping strength $t$, which
was also used as unit of energy, to $t=0.2 eV$, in accordance with
experiments, one obtains the Curie temperature in reasonable agreement
with experiment, see Fig. \ref{LaSrMnO_phases.eps}. In order to obtain
the different low temperature phases besides FM and PM observed in experiments for low
carrier concentrations, finite $J_H$ and $J'$ would be
needed as well as two orbitals with Coulomb repulsion.

\subsection{Phase Separation versus Ferromagnetic Polarons}
\label{sec:PS_FP}

The discontinuity of the filling as a function of the chemical potential, see
Fig. \ref{mu_n_L20Jx.eps}, is usually interpreted as phase separation
\cite{yunoki98:_phase}, i.e. the system is expected to split into
antiferromagnetic domains with low and ferromagnetic domains with
higher carrier concentration. Taking Coulomb interactions into account, PS has been
argued to lead to either small \cite{Sboychakov_02} or
large \cite{moreo_science_99} (nano-scale) clusters, which have been the
basis for a possible though controversial \cite{EdwardsI} explanation of
CMR \cite{dagotto01:review, Millis_III}. More thorough evaluation of
the MC data for the transition near the filled lower Kondo band
reveals however, that single hole ferromagnetic polarons 
are stabilized instead, even without any Coulomb repulsion. All results
presented in this section are for the effective spinless fermion model
\Eq{def:ESF}. 

\begin{figure}[h]
  \centering
  \subfigure{\includegraphics[width=0.3\columnwidth,height=0.4\columnwidth]
  {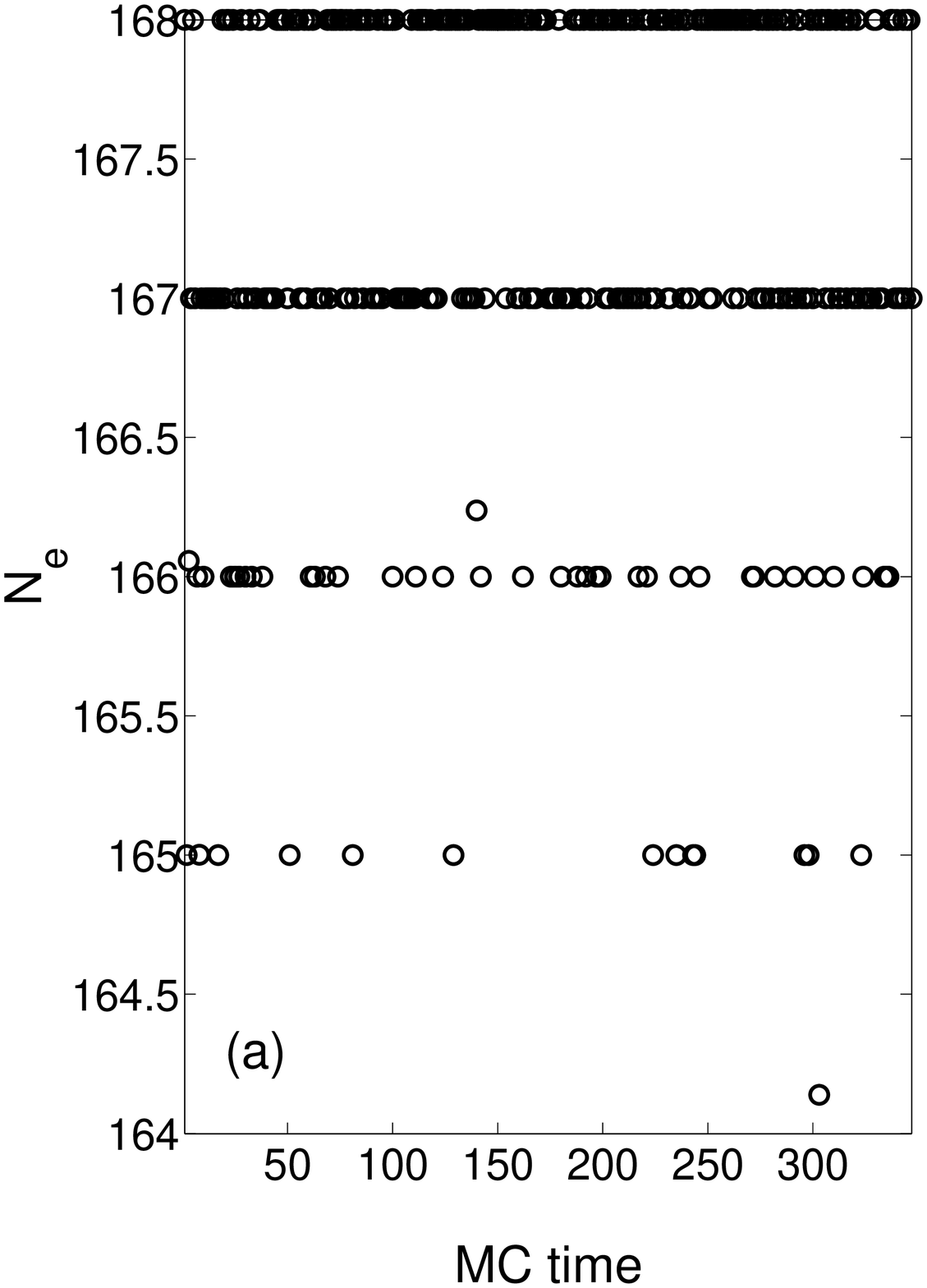}\label{mean_paricle_numbers_af}}
  \subfigure{\includegraphics[width=0.3\columnwidth,height=0.4\columnwidth]
  {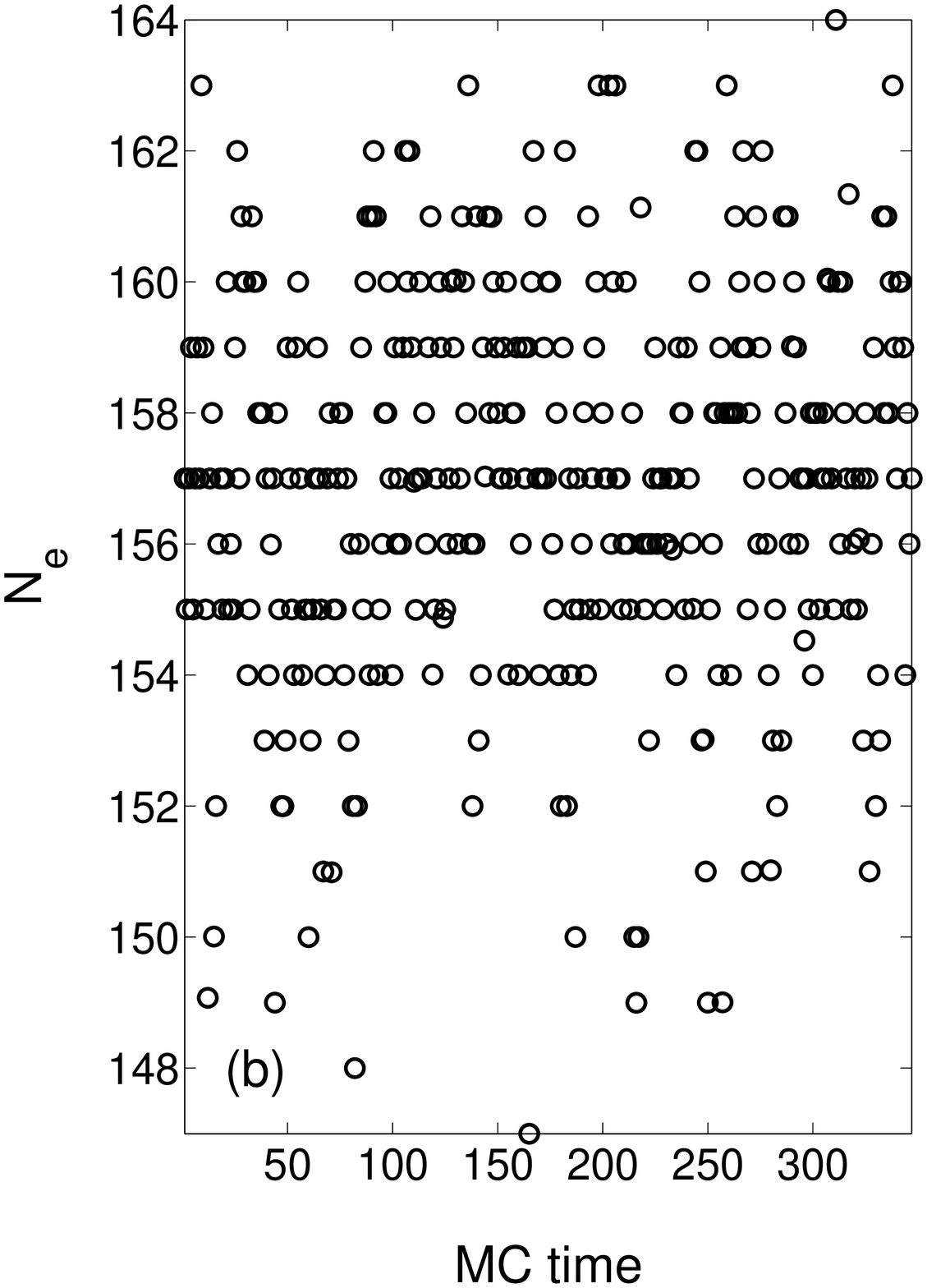}\label{mean_paricle_numbers_pol}}
  \subfigure{\includegraphics[width=0.3\columnwidth,height=0.4\columnwidth]
  {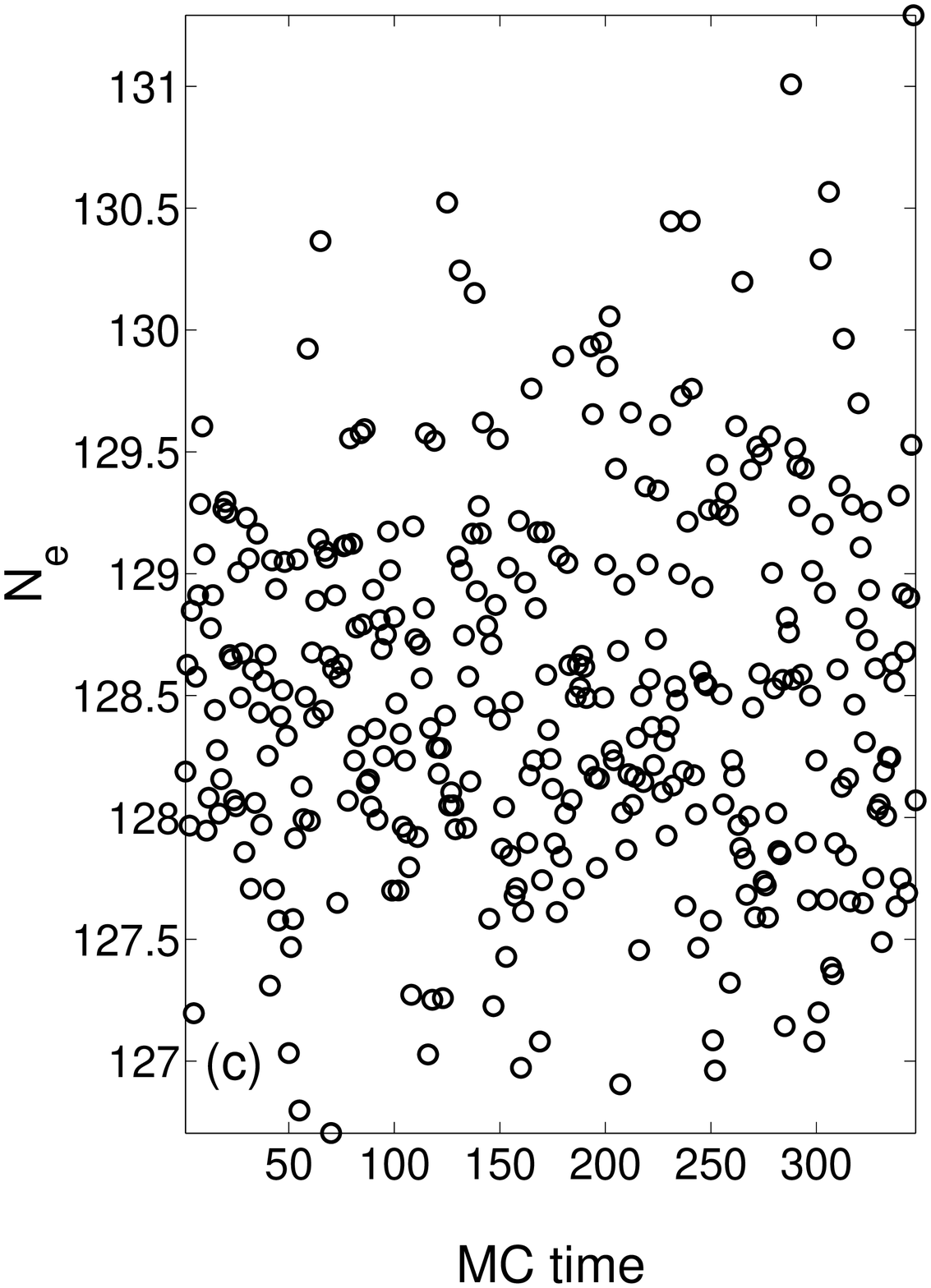}\label{mean_paricle_numbers_fm}}\\[-1em]
  \caption{Mean particle numbers $N_\textrm{e}$ in a grand canonical MC simulation in
  2 dimensions ($L=14\times 12,
    J_H=6, J'=0.02, \beta=50$) as a function of MC time. One time step
    corresponds to $200$ sweeps of the lattice.
    (a) $\mu=1.26>\mu^*$: almost filled,
    (b) $\mu=1.19 \simeq\mu^*$: polaron regime,
    (c) $\mu= 1.12 < \mu^*$: FM regime.
    For visibility, only the first 350 time steps are shown.
  \label{mean_paricle_numbers}}
\end{figure}

A first indication for ferromagnetic polarons is the behavior of the
electron density $\langle n \rangle_\mathcal{S}$
with the MC time near the critical chemical potential as depicted in
Fig. \ref{mean_paricle_numbers}, where $\langle n \rangle_\mathcal{S}$
is the thermodynamical expectation value for the filling given the
core-spin configuration $\mathcal{S}$ sampled by the MC run.
In the FM regime [Fig.\ref{mean_paricle_numbers_fm}], the
density fluctuates slightly
with the fluctuating core-spin configurations and takes
non-integer values in accordance with standard results for free
electrons. For the almost
filled band [Fig.\ref{mean_paricle_numbers_af}], the density is
$\langle n \rangle_\mathcal{S}=1$ (0 holes) for most spin configurations and 
occasionally, configurations occur, which contain exactly one, exactly two or exactly three holes. 
In between [Fig.\ref{mean_paricle_numbers_pol}], the particle number
fluctuates strongly, but as for $\langle n \rangle_\mathcal{S}\approx1$,
almost only integer fillings occur.
While these integer fillings can hardly be understood
in a PS scenario which is supposed to be a mixture of the low- and the
high-density phases, it
can easily be explained by independent 
polarons containing one hole each. As they are independent from each
other, all have the same energy exactly balanced by the critical
chemical potential $\mu^*$ and their number therefore fluctuates
strongly.

To measure the size of the ferromagnetic domains, we use the dressed
core-spin correlation function
\begin{equation}                                              \label{eq:hss}
  S_h(l)=\frac{1}{L-l}\,\sum_{i=1}^{L-l}
  n_i^h\;\vec{S}_i\cdot \vec{S}_{i+l}
%  S_h(\vec r) = \frac{1}{L} \sum_{\vec i} n^h_{\vec i} S_{\vec i} \cdot S_{\vec i + \vec r}
\end{equation}
that measures the correlations around a hole. The hole- density
operator $n_i^h$ is related to the electron density via $n_i^h = 1-n_i$.
Equation (\ref{eq:hss}) holds for open boundary conditions (employed
in 1D), the formula for periodic boundary conditions used in 2D
is analogous.

\begin{figure}[h]
  \begin{minipage}{0.45\textwidth}
    \includegraphics[width=\textwidth]{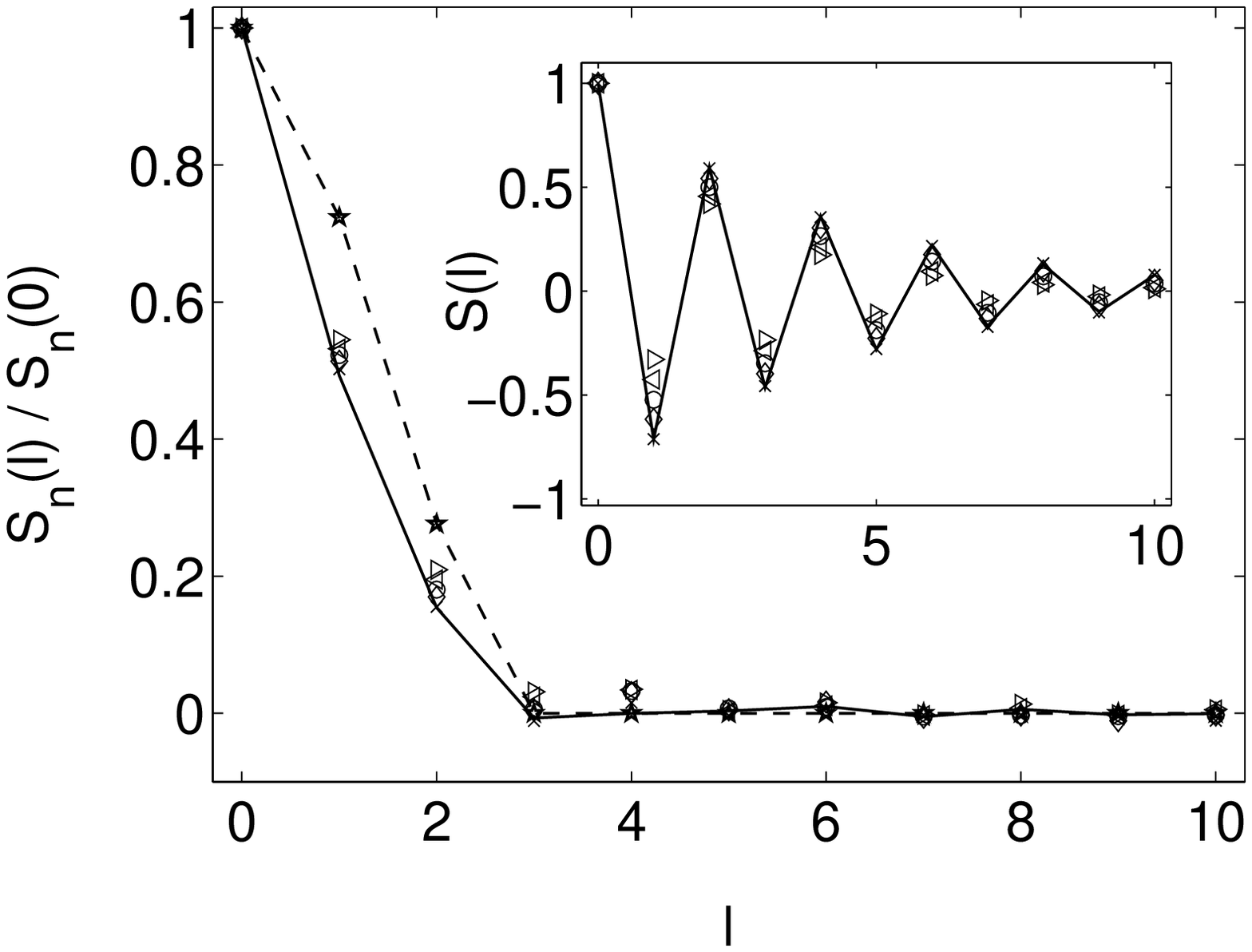}
  \end{minipage}\hfill
  \begin{minipage}{0.45\textwidth}
  \includegraphics[width = \textwidth]{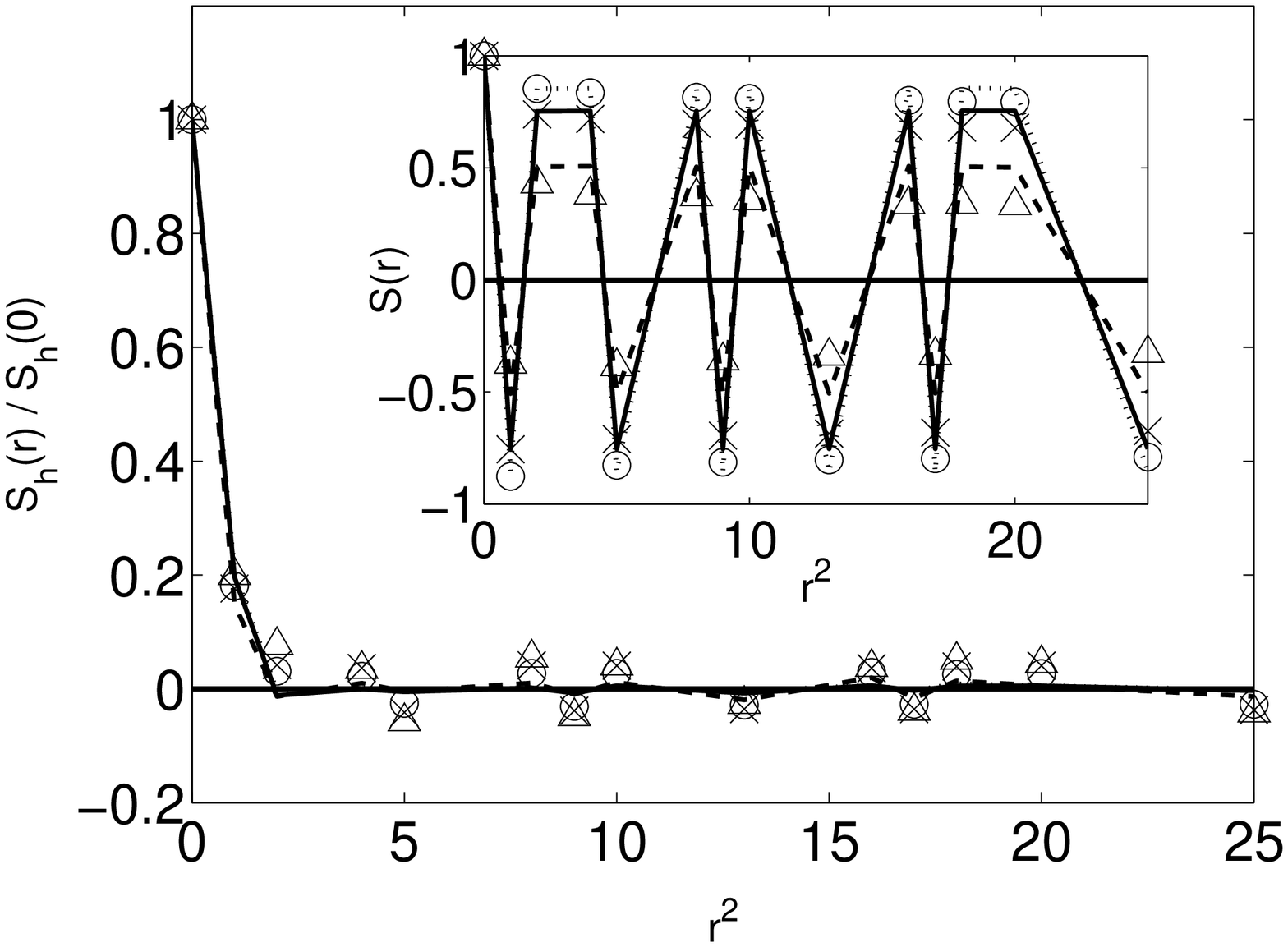}
  \end{minipage}
  \caption{Dressed spin-spin correlation function \Eq{eq:hss} from unbiased MC
    for $\beta=50$, $J'=0.02$, and $J_H=6$.
    Left panel: 1 dimension:
     $L=50$-site chain containing one ($\times$), two ($\diamond$), three
     ($\circ$), four $(\triangleleft)$, and five $(\triangleright)$ holes.
     The dashed line is calculated within the simple polaron picture, while
     the solid line represents the generalized UHA result for a single
    polaron, see \cite{KollerPruell2002c}.
     Right panel: 2 dimensions: $12 \times 14$ lattice with 1
     ($\circ$), 6 ($\times$) and 20 ($\triangle$) holes. Continuous lines are data
      for the simple Polaron model (see Fig. \ref{polaron.eps}, right panel): 1 Polaron
      (dotted), 6 (solid)
      and 20 Polarons (dashed).
      The insets show  the conventional spin-spin correlation function
     $  S(l)=\frac{1}{L-l}\,\sum_{i=1}^{L-l} \vec{S}_i\cdot
     \vec{S}_{i+l}$. Left from Ref.~\cite{KollerPruell2002c}, right from Ref.~\cite{DaghoferKoller2003}.
 }
  \label{css_polarons.eps}
\end{figure}

Figure \ref{css_polarons.eps} shows this dressed core-spin
correlation. The ferromagnetic regions around the holes are small and
their size
does not grow with doping neither for the one- nor for the the two-dimensional case. This
indicates that the introduction of more holes leads to more small FM
polarons rather than to a growth of the existing ones.

\begin{figure}
  \hfill  
  \begin{minipage}{0.49\textwidth}
    \vspace*{5mm}
    \psfrag{n}{\large $n_h$}
    \includegraphics[width=0.8\textwidth]{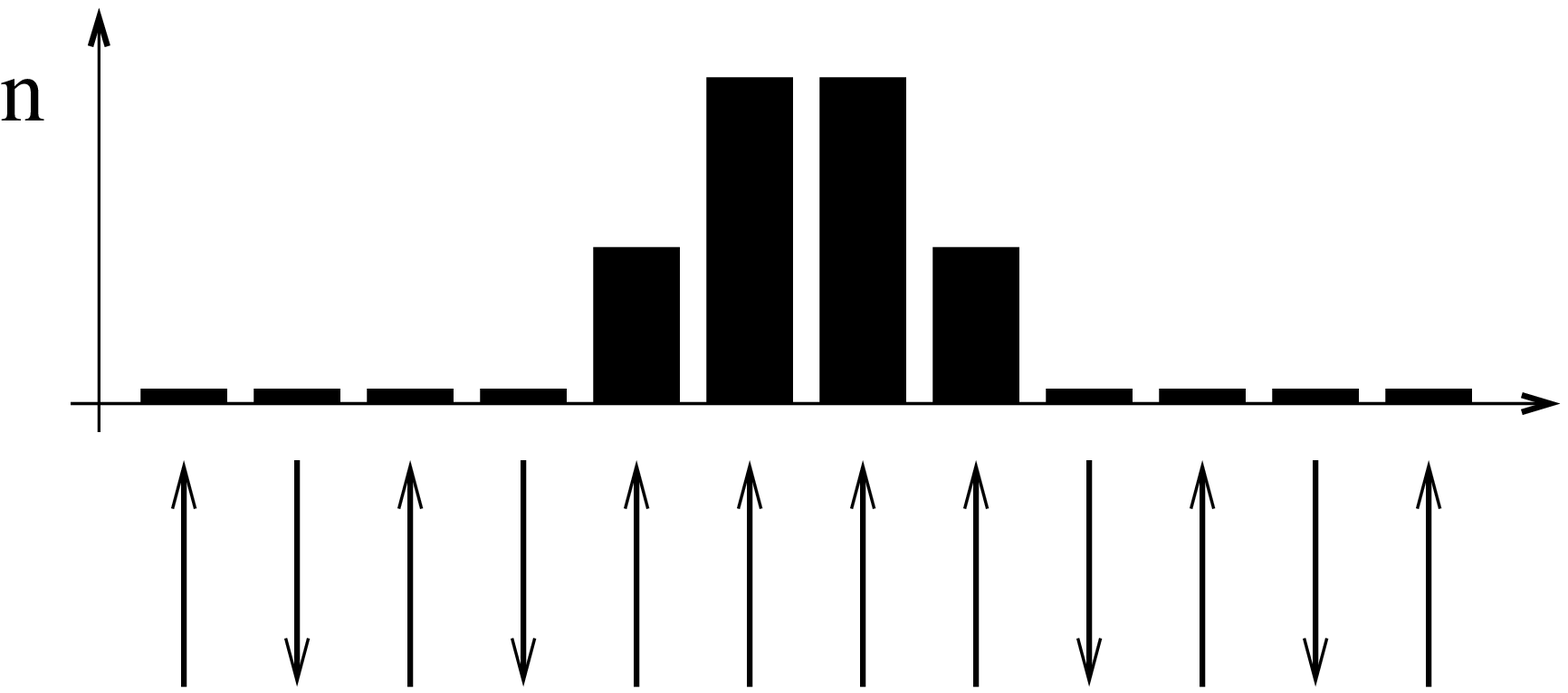}
  \end{minipage}\hfill
  \begin{minipage}{0.49\textwidth}
  \includegraphics[width=0.7\textwidth]{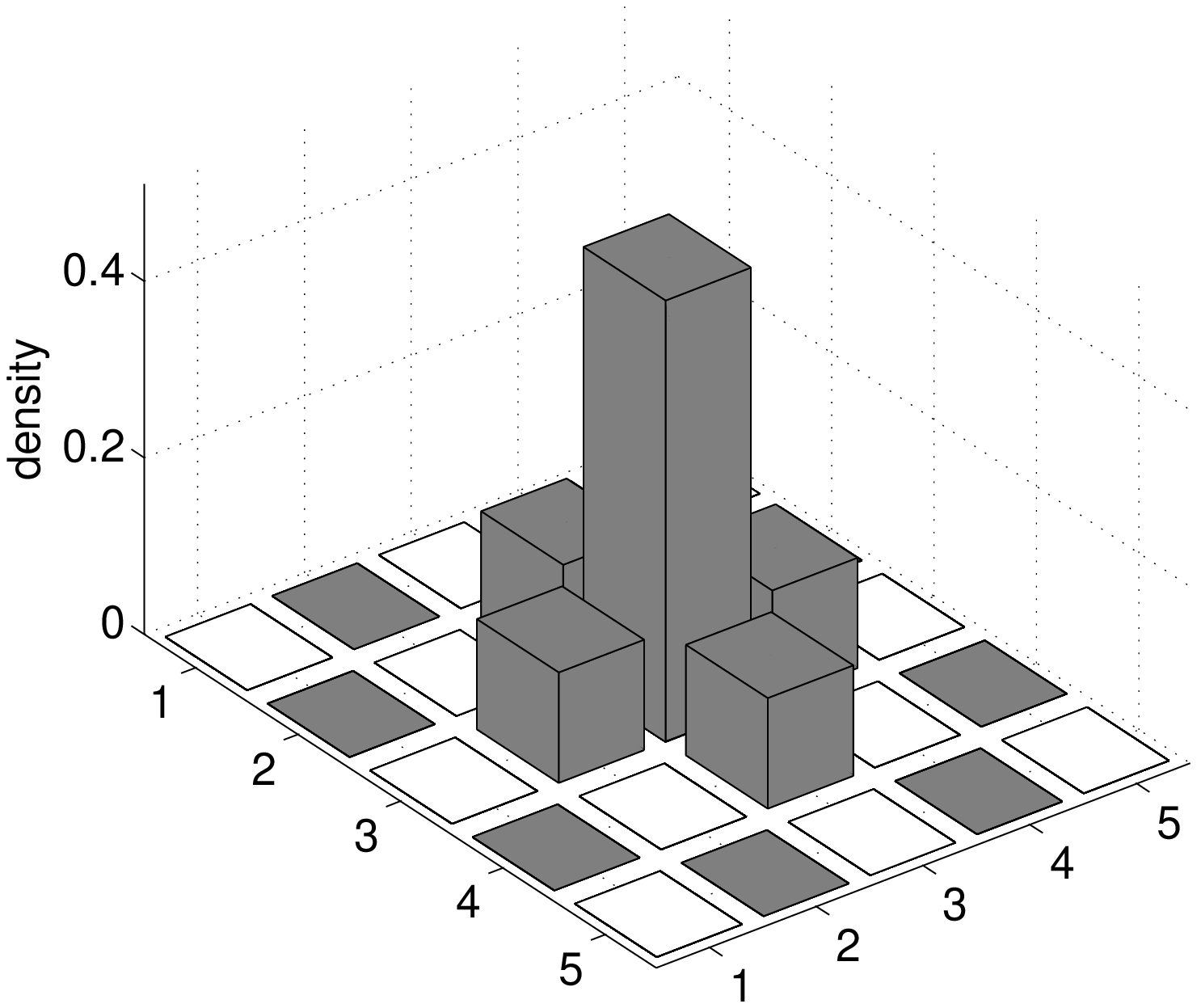}
  \end{minipage}
  \caption{Toy model for FM polarons in one (left) and two (right)
    dimensions. Height represents hole density.
    In 1 D, a FM domain of $L_f=4$ lattice sites is embedded in an AFM
    background, in 2D one spin is flipped from the perfect AFM.
    A single hole is localized in the FM domain giving rise to the depicted
    hole density (different from the schematic shape in Fig.~4 of
    Ref.~\cite{moreo99}). Left from Ref.~\cite{KollerPruell2002c}, right from Ref.~\cite{DaghoferKoller2003}.
  }
  \label{polaron.eps}
\end{figure}

These observations lead to the development of a simple toy model for
the FM polarons. Each polaron consist of a small (3-4 sites in 1D, 5
sites in 2D) FM well, in which the hole can delocalize. These wells
are embedded into an AFM background. For a schematic representation,
see Fig. \ref{polaron.eps}. The value of the critical chemical
potential can be obtained from the toy model by simple energy
considerations. It is simply the difference between the energy gained
by the delocalized hole and the energy payed for the breaking of AFM
bonds. In this simplest model, the hopping strength is given by $u_f=1$
for the FM regions and $u_a=0$ for AFM bonds. The impact of thermal
fluctuations of the core spins can be modeled by a generalization of
UHA to inhomogeneous structures, where $\Gamma(u_f, u_a)$ has to be determined. 

\begin{figure}[hbtp]
  \centering
   \includegraphics[width = 0.44\textwidth,height=0.4\columnwidth]{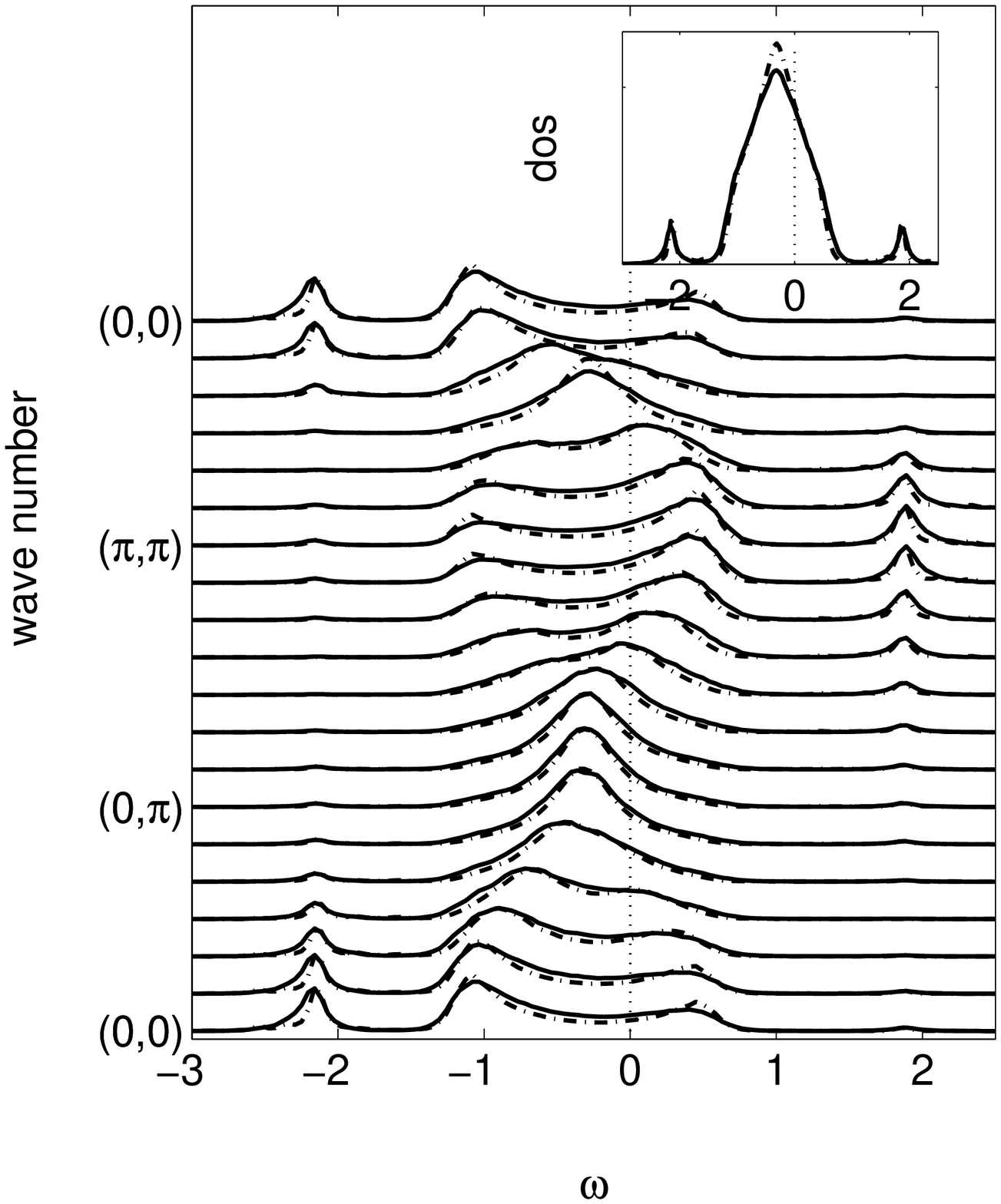}
   \includegraphics[width = 0.44\textwidth,height=0.4\columnwidth]{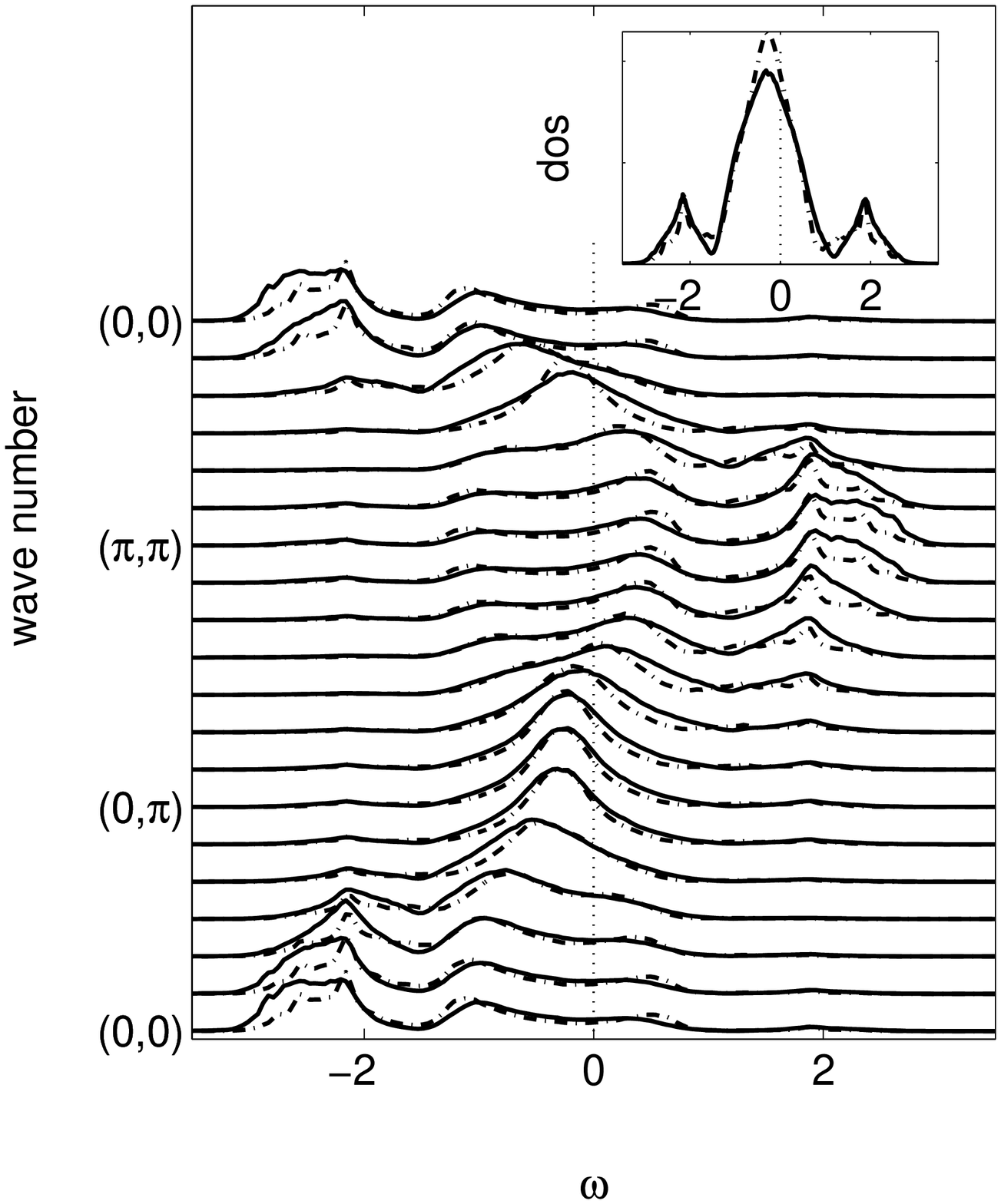}
   \caption{Spectral density for $J'=0.02, \beta=50$, $J_\textnormal{H}=6$ on a $12 \times 14$ lattice:
     left: $6$ holes ($x\approx 0.035$); right: $20$ holes ($x\approx
     0.12$). Solid lines are the unbiased MC data, dashed lines the
     simplified polaron model.}
  \label{fig:spec_Jse0.02_beta50_N162}
\end{figure}

This was done for the one-dimensional model, for results and details of the algorithm
see \cite{KollerPruell2002c}. In order to compare the toy model to the
MC data in 2 dimensions, random deviations
were added to the core spins, see \cite{DaghoferKoller2003}. The
principal effect of the fluctuations is a finite bandwidth for the AFM
at half filling, their amount was therefore fitted to yield the same width for
the AFM band as the MC data, see Fig. \ref{fig:spec_Jse0.02_beta50_N162}.

\begin{figure}[h]
  \psfrag{pi }{$\pi$}
  \psfrag{omega}{$\omega$}
  \includegraphics[width=0.32\columnwidth,height=0.35\columnwidth]{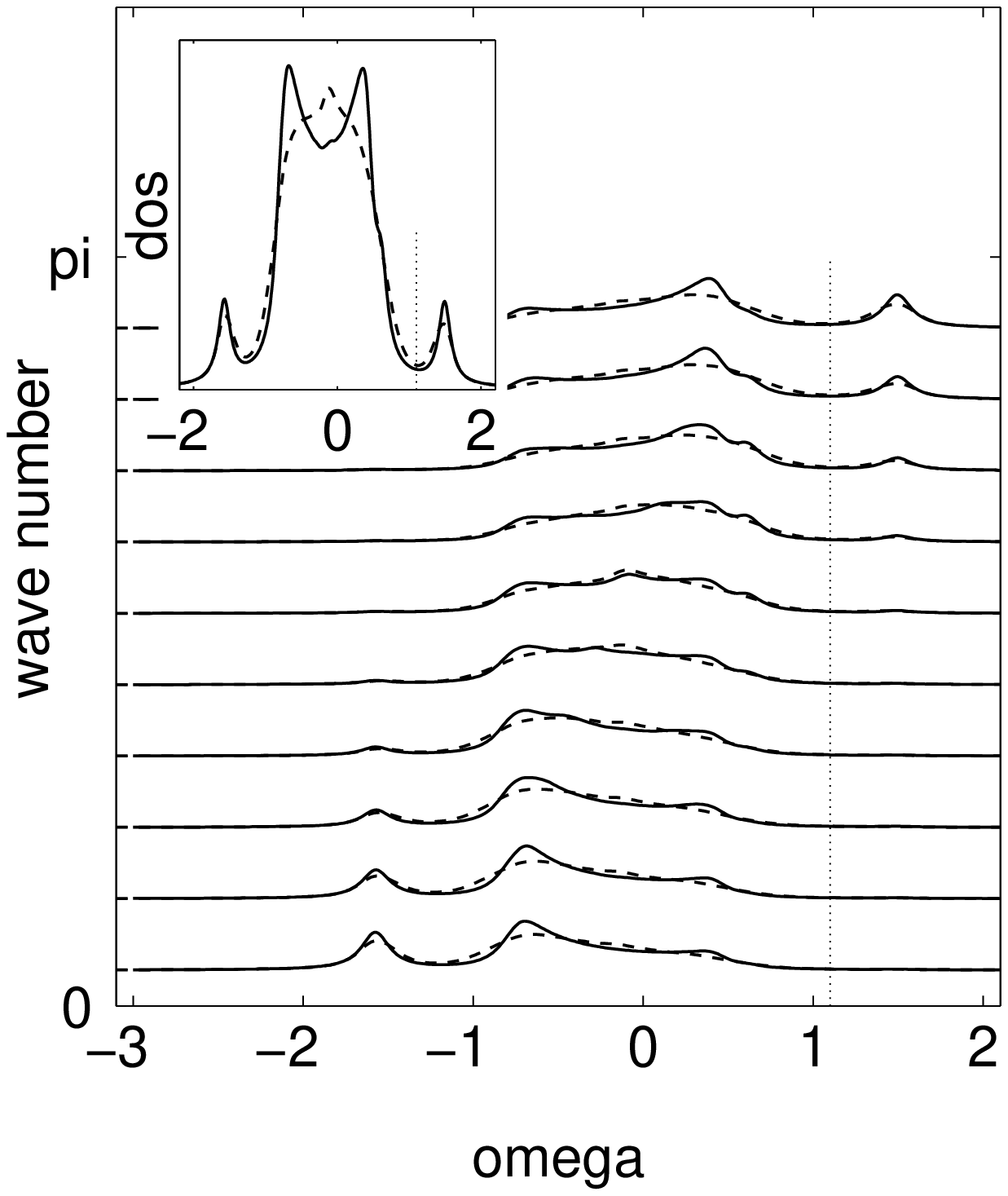} 
%  \hfill
  \includegraphics[width=0.32\columnwidth,height=0.35\columnwidth]
  {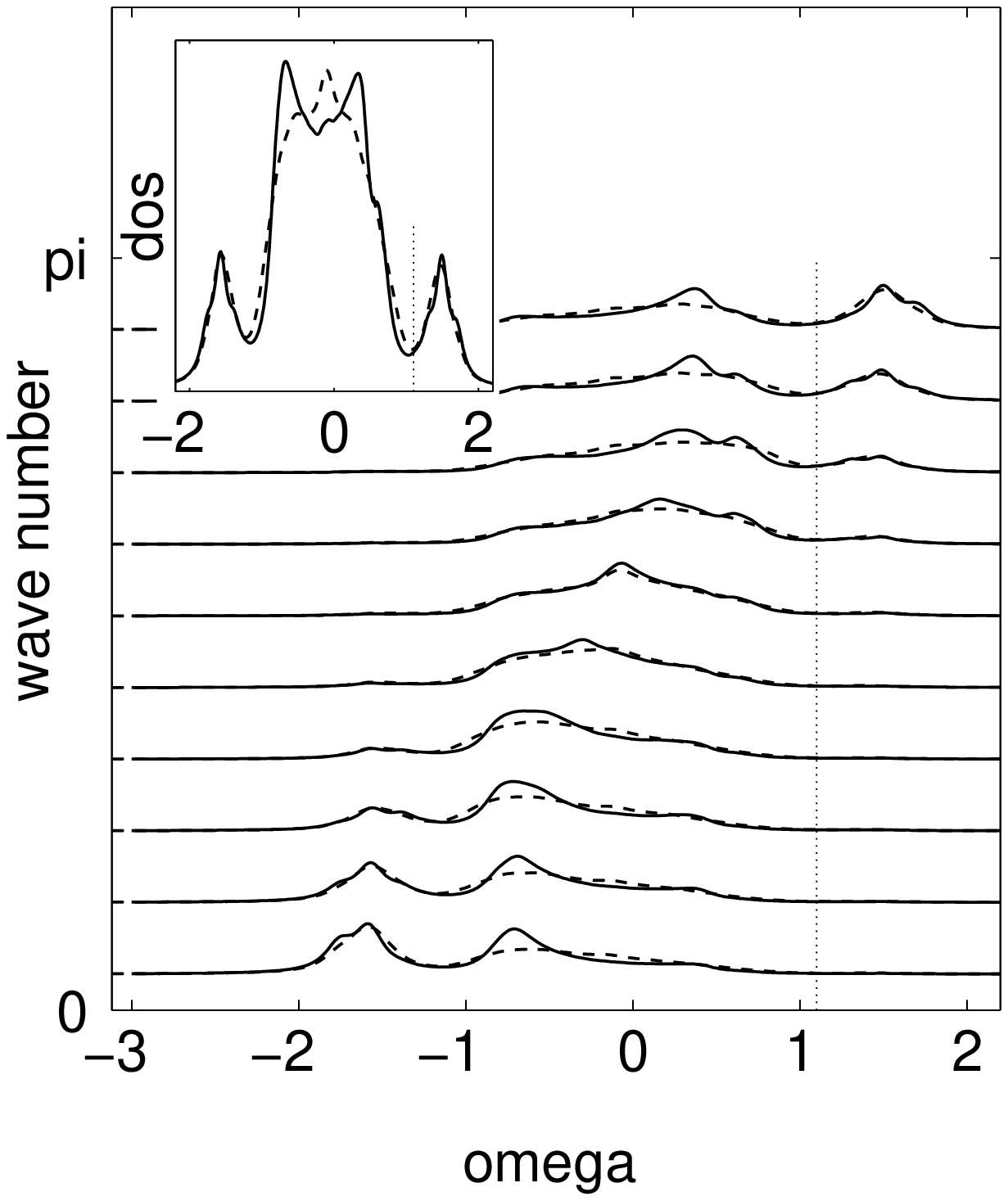}
  \hfil
  \includegraphics[width=0.32\columnwidth,height=0.35\columnwidth]
  {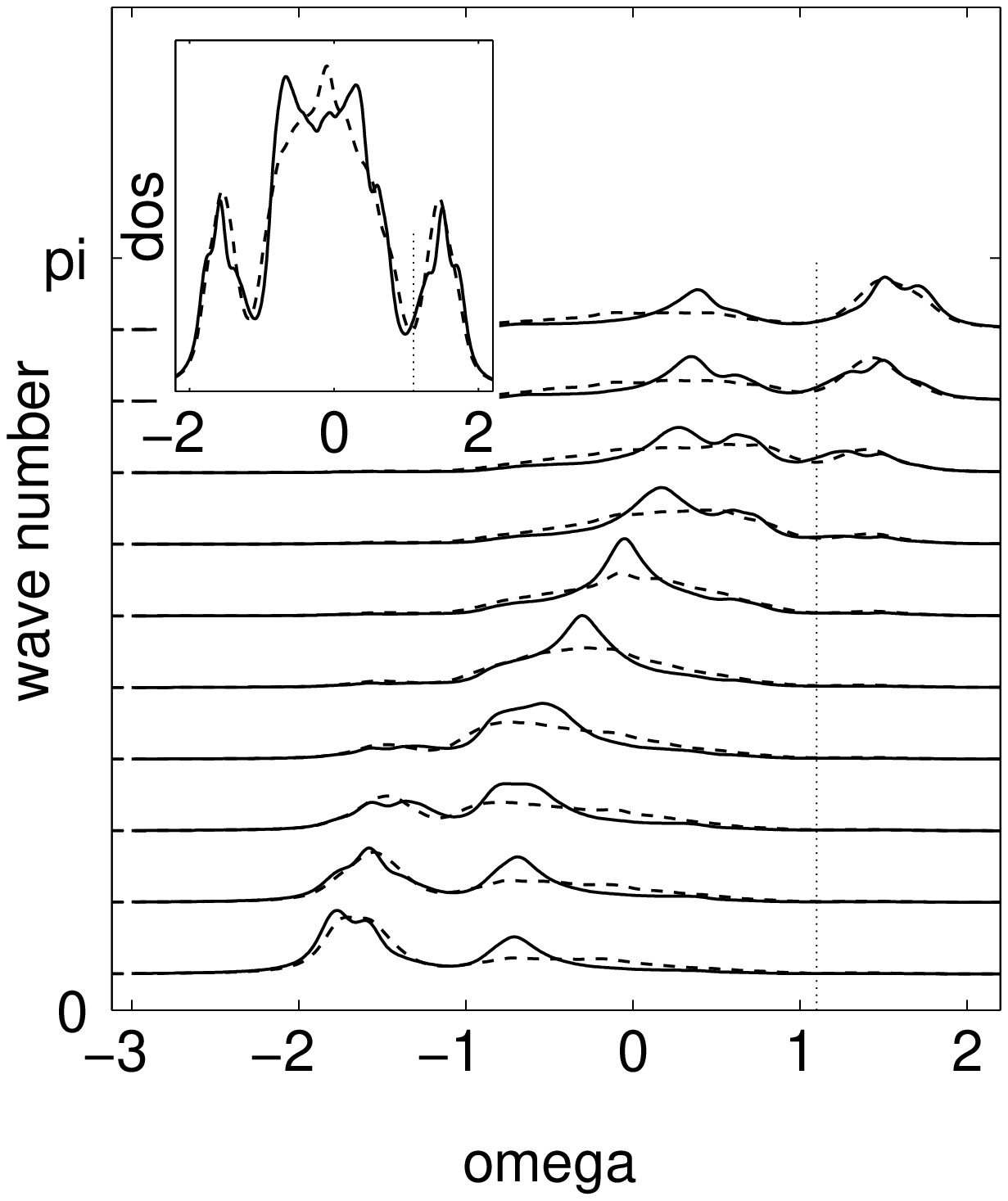}
  \caption{Spectral density for $N_h=1$ hole, $N_h=2$ holes $N_h=3$
  holes, corresponding to 1, 2 and 3 polarons in one dimension. Dashed
    lines: MC data, solid: generalized UHA.
    Parameters as in Fig.~\ref{mean_paricle_numbers}. From Ref.~\cite{KollerPruell2002c}.}
  \label{sw_L20N18_beta50.eps}
\end{figure}

Because the FM wells in which the hole can
move are so small, they give rise to only a few well separated signals
in the spectral density. Figure \ref{sw_L20N18_beta50.eps} shows the
spectral density and the density of states for one, two and three
holes in one dimension. On sees a broad band in the center which comes
from holes moving in the imperfect AFM background. Separated from this central
band by a (mirror) pseudogap are dispersionless states from the FM polarons at
$\omega \simeq \pm 1.5$. The weights of these signals increase upon
the introduction of more holes.

Figure \ref{fig:spec_Jse0.02_beta50_N162} shows the spectral density
for the two dimensional model with 6 and 20 holes and compares it to
the data for the toy model with added random fluctuations. There is
again a broad central band from the AFM featuring a mirror band due to
the doubling of the unit cell and again the polaronic states
separated by a pseudogap. There is strong correspondence of the
unbiased MC data to the simplified model for both the one- and the two dimensional system.

The pseudogap, which is also observed in experiments \cite{DessauI,
DessauII, DessauIII,Park95}, can easily be explained by the few well separated
eigenenergies of the holes trapped in the small polarons. In a phase separation scenario
with  larger FM regions, additional states would fill the gap between
the AFM band and the polaron states, in contrast to experimental
results and the MC data.

\subsection{Phase diagram in 2D}
\label{sec:phase_diagr}

Although $0<J'<0.1$ is by far the smallest parameter of the
Hamiltonian, it has a considerable effect; especially at low carrier
density, when the kinetic energy is small. While we observed nearly
independent polarons for $\beta=50$ and $J'=0.02$, they tend to form
larger clusters and eventually phase separation for decreasing $J'$ and avoid
each other for larger $J'$. The reason for this effect is the
stabilization of the AFM background by $J'$.

\begin{figure}[ht]
  \centering
  \includegraphics[width=0.55\textwidth,height=0.4\textwidth]{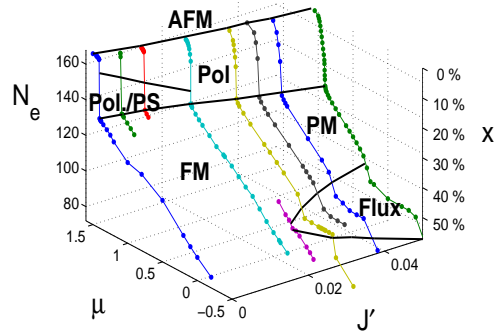}
  \caption{Electron number $N_e$ as a function of $\mu$ and $J'$, and phase
    diagram for $-0.5< \mu < 1.6$, $0 \leq J' \leq 0.05$, $0 \leq x \leq 0.6$ (i.~e. filled to $40\%$ filled lower Kondo
    band),$J_\textnormal{H} = 6, \beta = 50$ on a $14 \times 12$-lattice. ``Pol.'': polaronic regime,
  ``Pol./PS'': mixture of both polarons and larger ferromagnetic
  clusters, ``FM'': ferromagnet, ``AFM'': antiferromagnet, ``PM'': regime without magnetic
  structure, ``Flux'': Flux phase.} 
\label{fig:phase_diagram}
\end{figure}

In order to determine the phase boundary between the polaronic regime
and phase separation, we chose the filling, at which the nearest AFM
signal (at the distance $r^2=5$) in the dressed core-spin correlation \Eq{eq:hss} became
ferromagnetic. This criterion is somewhat arbitrary and the
transition is not sharp,  polarons rather coexisting with larger clusters.

For large doping, $J'=0.05$ destroys ferromagnetism and leads to the so
called flux phase around half filling~\cite{Aliaga_island_2d,Agterberg_00,Yamanaka_98,DaghoferSCES}.
In the phase diagram, one sees a small window in parameter space for
phase separation, but for realistic parameters $J' > 0.01$, polarons
dominate, especially at small doping. 

\section{Summary}
\label{sec:sum}
 
An effective spinless fermion model (ESF) was derived from the FM Kondo
lattice model \Eq{def:FKLM} with classical core spins and for $J_H\gg t$,
leading to the ESF Hamiltonian \Eq{def:ESF}. It allows the treatment of finite $J_H$ with the 
same numerical effort as the $J_H \to \infty$ approximation, but
gives better correspondence to the original model. A further
simplification was achieved by treating the fluctuating
core spins by a uniform hopping strength (UHA). With this much
simpler model \Eq{UHA_Ham}, we obtained the Curie temperature for the 3D
model with one itinerant orbital in accordance with experimental values.

By unbiased MC simulations for the ESF
model \Eq{def:ESF} with non-degenerate conduction band in one and two dimensions, we found that ferromagnetic polarons
are the reason for features previously attributed to phase
separation. This polaronic behavior is enhanced by larger $J' >
0.02$. A phase diagram was obtained for the 2 dimensional case.

\printindex
\end{document}